\documentclass[sigconf]{acmart}

\pdfoutput=1

\usepackage{graphicx}
\usepackage{balance}  

\usepackage{color}
\usepackage{url}

\usepackage[ruled, noend]{algorithm2e}

\usepackage{tikz}
\usetikzlibrary{fit,topaths,calc,shapes,snakes,chains,scopes}
\usepackage{subfigure}
\definecolor{goodgreen}{rgb}{0.1, 0.5, 0.1}

\usepackage[normalem]{ulem}
\usepackage{soul}

\newcommand{\pluseq}{\mathrel{+}=}
\newcommand{\TAB}{\makebox[2.5ex][r]{}}%
\newcommand{\STAB}{\makebox[1.5ex][r]{}}%
\newcommand{\IF}{\textbf{if}\xspace}%
\newcommand{\FOREACH}{\textbf{foreach}\xspace}%
\newcommand{\DO}{\textbf{do}\xspace}%
\newcommand{\RETURN}{\textbf{return}\xspace}%
\newcommand{\MATCH}{\textbf{switch}\xspace}%
\newcommand{\SUM}{\texttt{SUM}\xspace}%
\newcommand{\COUNT}{\texttt{COUNT}\xspace}%

\newcommand{\aggs}{\mathsf{aggregates}}

\newcommand{\rfbox}[1]{{\color{red}\fbox{\color{black}#1}}}

\newcommand{\fama}{\text{\sf FaMa}}
\newcommand{\pr}{\text{\sf PR}}
\newcommand{\lr}{\text{\sf LR}}

\newcommand{\R}{\mathbb R}

\newcommand{\nop}[1]{}

\newcommand{\inner}[1]{\left\langle #1 \right\rangle}

\newcommand{\mv}[1]{\mathbf{#1}}
\renewcommand{\vec}[1]{\ensuremath\boldsymbol{#1}}

\newcommand{\norm}[1]{\left\|#1\right\|}
\newcommand{\gv}[1]{\ensuremath{\mbox{\boldmath$ #1 $}}} 
\newcommand{\grad}[1]{\gv{\nabla} #1} 
\newcommand{\pd}[2]{\frac{\partial#1}{\partial#2}}

\DeclareMathOperator*{\argmin}{arg\,min}

\usepackage{todonotes}

\allowdisplaybreaks[1]

\setcopyright{none}
\renewcommand\footnotetextcopyrightpermission[1]{} 
\begin{document}

\title{AC/DC: In-Database Learning Thunderstruck}

\author{Mahmoud Abo Khamis}
\affiliation{\institution{RelationalAI, Inc}}
\author{Hung Q. Ngo}
\affiliation{\institution{RelationalAI, Inc}}
\author{XuanLong Nguyen}
\affiliation{\institution{University of Michigan}}
\author{Dan Olteanu}
\affiliation{\institution{University of Oxford}}
\author{Maximilian Schleich}
\affiliation{\institution{University of Oxford}}

\renewcommand{\shortauthors}{M. Abo Khamis, H.Q. Ngo, X. Nguyen, D. Olteanu, M. Schleich}

\copyrightyear{2018} 
\acmYear{2018} 
\setcopyright{acmlicensed}
\acmConference[DEEM'18]{International Workshop on Data Management for 
End-to-End Machine Learning}{June 15, 2018}{Houston, TX, USA}
\acmBooktitle{DEEM'18: International Workshop on Data Management for 
End-to-End Machine Learning, June 15, 2018, Houston, TX, USA}
\acmPrice{15.00}
\acmDOI{10.1145/3209889.3209896}
\acmISBN{978-1-4503-5828-6/18/06}

\begin{abstract}
We report on the design and implementation of the AC/DC gradient descent solver 
for a class of optimization problems over normalized databases. AC/DC 
decomposes an optimization problem into a set of aggregates over the join 
   of the database relations. It then uses the answers to these aggregates to 
iteratively improve the solution to the problem until it converges.

The challenges faced by AC/DC are   
the large database size, 
the mixture of continuous and categorical features, and 
the large number of aggregates to compute. 
AC/DC addresses these challenges by employing 
a sparse data representation, 
factorized computation, 
problem reparameterization under functional dependencies, and 
a data structure that supports shared computation of aggregates.

To train polynomial regression models and factorization machines of up to 154K features over the natural join of all relations from a real-world dataset of up to 86M tuples, AC/DC needs up to 30 minutes on one core of a commodity machine. This is up to three orders of magnitude faster than its competitors R, MadLib, libFM, and TensorFlow whenever they finish and thus do not exceed memory limitation, 24-hour timeout, or internal design limitations. 

\end{abstract}

\maketitle

\begin{quote}
{\em 
\hspace*{-1em}Rode down the highway\\
\hspace*{-1em}Broke the limit, we hit the town\\
\hspace*{-1em}Went through to Texas, yeah Texas, and we had some fun.\\
\hspace*{12em} -- Thunderstruck (AC/DC)
}
\end{quote}

\section{Introduction}
\label{sec:introduction}

In this paper we report our on-going work on the design and implementation of
AC/DC, a gradient descent solver for a class of optimization problems including
ridge linear regression, polynomial regression, and factorization machines.  It
extends our prior system F for factorized learning of linear regression
models~\cite{SOC16} to capture non-linear models, categorical features, and
model reparameterization under functional dependencies (FDs).  Its design is but
one fruit of our exploration of the design space for the AI engine currently
under development at RelationalAI.  It subscribes to a recent effort to
bring analytics inside the
database~\cite{MADlib:2012,Kumar:InDBMS:2012,KuNaPa15,SOC16,SPOOF:CIDR:2017} and
thereby avoid the non-trivial time spent on data import/export at the interface
between database systems and statistical packages.

AC/DC\footnote{AC/DC supports both categorical and
  continuous features and fast processing. Its name allures at the duality of
  alternating and discrete currents and at the fast-paced sound of a homonymous
  rock band.} solves optimization problems over design matrices defined by
feature extraction queries over databases of possibly many relations. It is a
unified approach for computing both the optimizations and the underlying
database queries; the two tasks not only live in the same process space, they
are intertwined in one execution plan with asymptotically lower complexity than
that of either one in isolation. This is possible due to several key
contributions.

First, AC/DC {\em decomposes} a given optimization problem into a set of
aggregates whose answers are fed into a gradient descent solver that iteratively
approximates the solution to the problem until it reaches convergence. The
aggregates capture the combinations of features in the input data, as required
for computing the gradients of an objective function.  They are group-by
aggregates in case of combinations with at least one categorical feature and
plain scalars for combinations of continuous features only. The former
aggregates are grouped by the query variables with a categorical domain.  Prior
work on in-database machine learning mostly considered continuous features and
one-hot encoded categorical features,
e.g.,~\cite{Kumar:InDBMS:2012,KuNaPa15,SOC16,SPOOF:CIDR:2017}. We avoid the
expensive one-hot encoding of categorical features by a {\em sparse
  representation using group-by aggregates}~\cite{NNOS2018}.  Several tools,
e.g., libFM ~\cite{libfm,liblinear} for factorization machines and
LIBSVM~\cite{CC01a} for support vector machines, employ sparse data
representations that avoid the redundancy introduced by one-hot encoding. These
are computed on the result of the feature extraction query once it is exported
out of the database.

Second, AC/DC {\em factorizes} the computation of these aggregates over the
feature extraction query to achieve the lowest known complexity.  We recently
pinpointed the complexity of AC/DC~\cite{NNOS2018}.  The factorized computation
of the queries obtained by decomposing optimization problems can be
asymptotically faster than the computation of the underlying join query alone.
This means that all machine learning approaches that work on a design matrix
defined by the result of the database join are asymptotically suboptimal.  The
only other end-to-end in-database learning system~\cite{KuNaPa15} that may outperform the
underlying database join works for generalized linear models over
key-foreign key joins, does not decompose the task to AC/DC's granularity, and
cannot recover the good complexity of AC/DC since it does not employ factorized
computation.

Third, AC/DC {\em massively shares computation} across the aggregate-join
queries. These {\em different} queries use the same underlying join and their
aggregates have similar structures.

Fourth, AC/DC {\em exploits functional dependencies} (FDs) in the input database
to reduce the dimensionality of the optimization problem. Prior work exploited
FDs for Na\"ive Bayes classification and feature
selection~\cite{Kumar:SIGMOD:16}. AC/DC can reparameterize (non)linear
regression models with non-linear regularizers~\cite{NNOS2018}. This
reparameterization requires taking the inverses of matrices, which are sums of
identity matrices and vector dot products.  To achieve performance improvements
by model reparameterization, AC/DC uses an interplay of its own data structure
and the Eigen library for linear algebra~\cite{eigenweb}.
  
  In this paper we report on the performance of AC/DC against MADlib~\cite{MADlib:2012},
  R~\cite{R-project}, libFM~\cite{libfm}, TensorFlow\cite{tensorflow} and our
  earlier prototype F~\cite{SOC16}. We used a real-world dataset with five
  relations of 86M tuples in total and up to 154K features to train a ridge
  linear regression model, a polynomial regression model of degree two, and a
  factorization machine of degree two. AC/DC is the fastest system in our
  experiments. It is orders of magnitude faster than the competitors or can finish
  successfully when the others exceed memory limitation, 24-hour timeout, or
  internal design limitations.

  The performance gap is attributed to the optimizations of AC/DC, which none of
  our competitors support fully. TensorFlow, R, and libFM require the
  materialization of the feature extraction query, as well as exporting/importing
  the query result from the database system to their system. They also require a
  transformation of the data into a sparse representation of the one-hot encoded
  training data before learning. MADlib does not materialize and export the query,
  but still requires an upfront one-hot encoding of the input relations, which comes
  with higher asymptotic complexity and prohibitively large relations with lots of
  zero entries. None of these systems benefit from factorized computation nor
  exploit functional dependencies. F is designed for linear regression models. 
  It uses factorized and shared computation of aggregates. It however does
  not exploit functional dependencies and requires the same one-hot encoding of
  categorical features in the input relations as MADlib.

  Our results confirm a counter-intuitive theoretical result from~\cite{NNOS2018}
  stating that, under certain conditions, exploiting query structures and smart
  aggregation algorithms, one can train a model using batch gradient descent (BGD)
  {\em faster} than scanning through the data once. In particular, this means BGD
  can be faster than one epoch of stochastic gradient descent (SGD), in contrast
  to the commonly accepted assumption that SGD is typically faster than BGD.

The paper is organized as follows. Section~\ref{sec:problem} presents the class
of optimization problems supported by AC/DC. Section~\ref{sec:theory} overviews
the foundations of AC/DC. Section~\ref{sec:evaluation} describes the data
structures used by AC/DC, its optimizations for factorized and shared
computation of aggregates. Section~\ref{sec:gradient} discusses the
reparameterization of optimization problems and their computation using the
aggregates computed in a previous step. Section~\ref{sec:experiments} reports on
experiments.

\section{Optimization Problems}
\label{sec:problem}

  We consider solving an optimization problem of a particular form
  inside a database; a more general problem formulation is presented
  in~\cite{NNOS2018}. Suppose we have $p$ parameters
  $\vec\theta=(\theta_1,\dots,\theta_p)$ and training dataset $T$ that contains
  tuples with $n$ features $\mv x = (x_1, \dots, x_n)$ and response $y$.  The
  features come in two flavors: continuous, e.g., {\sf price}, and
  qualitative/categorical, e.g., {\sf city}. The former are encoded as scalars,
  e.g., the price is 10.5. The latter are one-hot encoded as indicator
  vectors, e.g., if there were three cities then the vector $[1,0,0]$ indicates
  that the first city appears in the training record.

  The ``learning'' phase in machine learning typically comes down to solving the
  optimization problem $\vec\theta^* := \argmin_{\vec\theta} J(\vec\theta)$,
  where $J(\vec\theta)$ is the loss function. We use the square loss and
  $\ell_2$-regularizer:
  \begin{equation}
    J(\vec\theta) = \frac{1}{2|T|}\sum_{(\mv x,y)\in T}
    \left(\inner{g(\vec\theta), h(\mv x)} - y\right)^2 +
    \frac \lambda 2 \norm{\vec\theta}_2^2.
    \label{eqn:general:form:of:J}
  \end{equation}
  The model is $\inner{g(\vec\theta), h(\mv x)}$, where $g$ and $h$ are
  parameter-mapping, and respectively feature-mapping, functions that uniformly
  capture continuous and categorical variables. In the case of continuous
  features only, $g$ and $h$ are vector-valued functions $g : \R^p \to \R^m$ and
  $h : \R^n \to \R^m$, where $m$ is a positive integer. Each component function
  $g_j$ of $g = (g_j)_{j\in [m]}$ is a multivariate {\em polynomial}.  Each
  component function $h_j$ of $h = (h_j)_{j\in [m]}$ is a multivariate {\em
    monomial}.  If there are categorical variables as well, the components of
  $h$ and $g$ become tensors. See~\cite{NNOS2018} and Example~\ref{ex:lr} below for details.

  The training dataset $T$ is the result of a feature extraction query $Q$ over an input database $D$ consisting of several relations. It is common for $Q$ to be the natural join of the relations in $D$ and select the columns hosting the desired features. Furthermore, $Q$ can be enhanced with additional selections and aggregates to construct new (intensional) features based on the input (extensional) ones.
    
  We next show how Eq.~\eqref{eqn:general:form:of:J} captures various regression models.

\begin{example}\label{ex:lr}
  Consider a feature extraction query that returns tuples over the variables
  $\{units\_sold, \textsf{city, country}, price\}$, where $units\_sold$, $price$
  are continuous and $\textsf{city, country}$ are categorical.
  
  A {\em ridge linear regression } ($\lr$) model with response $units\_sold$ and
  features $\{\textsf{city, country}, price\}$ can be learned by
  optimizing~\eqref{eqn:general:form:of:J} with the functions
  $\mv g(\vec\theta)=(\theta_0, \vec\theta_{\sf city}, \vec\theta_{\sf country},
  \theta_{price})$ and
  $h(\mv x) = (1, \mv x_{\sf city}, \mv x_{\sf country}, x_{price})$. The
  entries in $g$ and $h$ for ${\sf city}$ and ${\sf country}$ are vectors,
  because the features $\mv x_{\sf city}, \mv x_{\sf country}$ are indicator vectors. 

  To learn a {\em degree-$2$ polynomial regression} ($\pr_2$) model, we extend
  the function $h$ from the $\lr$ model with all pairwise combinations of
  features: $\mv x_{\sf city} \otimes\mv x_{\sf country}$,
  $\mv x_{\sf city} x_{price}$, $\mv x_{\sf country} x_{price}$, and
  $x_{price}^2$. We do not include $\mv x_{\sf city}\otimes\mv x_{\sf city}$ and
  $\mv x_{\sf country}\otimes\mv x_{\sf country}$ because they encode the same
  information as the original indicator vectors. The function $g$ from the $\lr$
  model is correspondingly extended with parameters for each interaction, e.g.,
  the parameter vector  $\vec \theta_{({\sf city}, price)}$ for the vector of interactions
  $\mv x_{\sf city} x_{price}$.

  Similarly, the {\em degree-$2$ rank-$r$ factorization machines} ($\fama_2^r$)
  model can be learned by extending the $h$ function from $\lr$ with all
  pairwise interactions of {\em distinct} features as for $\pr_2$, yet
  without the interaction $x^2_{price}$. In contrast to $\pr_2$,
  the parameters corresponding to
  interactions are now factorized: The entry in $g$ corresponding to the
  interaction $\mv x_{\sf city} x_{price}$ is
  $\sum_{\ell=1}^r \vec \theta_{\sf city}^{(\ell)}
  \cdot\theta_{price}^{(\ell)}$.
\end{example}

\section{Overview of AC/DC Foundations}
\label{sec:theory}

AC/DC is a batch gradient-descent (BGD) solver that optimizes
the objective $J(\vec\theta)$ 
over the training dataset $T$ defined by a
feature extraction query $Q$ over a database $D$. Its high-level structure is
given in Algorithm~\ref{algo:bgd}. 
The inner loop repeatedly computes the loss function $J(\vec\theta)$ and its gradient $\grad J(\vec\theta)$. This
can be sped up massively by factoring out the data-dependent computation 
from the optimization loop~\cite{SOC16,NNOS2018}. The former
is cast as computing aggregates over joins, which benefit from recent
algorithmic advances~\cite{faq,fdb}.

\begin{algorithm}[h]
   \caption{BGD with Armijo line search.}
   \label{algo:bgd}
   $\vec\theta \gets $ a random point\;
   \While{not converged yet}{
      $\alpha \gets $ next step size \tcp*[f]{Barzilai-Borwein~\cite{MR967848}}\;
      $\mv d \gets \grad J(\vec\theta)$\;
      \While{$\left(J(\vec\theta-\alpha \mv d) \geq J(\vec\theta)-\frac
      \alpha 2 \norm{\mv d}_2^2\right)$}{
         $\alpha \gets \alpha/2$ \tcp*[f]{line search}\;
      }
      $\vec\theta \gets \vec\theta - \alpha \mv d$\;
   }
\end{algorithm}

{\noindent\bf From Optimization to Aggregates.} Let us define the matrix $\vec\Sigma = (\vec\sigma_{ij})_{i,j\in [m]}$, 
the vector $\mv c = (\vec c_i)_{i \in [m]}$, and the scalar $s_Y$ by
\begin{eqnarray}
   \vec\sigma_{ij} &=& \frac{1}{|Q(D)|} \sum_{(\mv x,y)\in Q(D)} h_i(\mv x)\otimes h_j(\mv
   x) \label{eqn:sigma}\\
   \vec c_i &=& \frac{1}{|Q(D)|} \sum_{(\mv x,y)\in Q(D)} y\cdot h_i(\mv x)\label{eqn:c}\\
   s_Y &=& \frac{1}{|Q(D)|}\sum_{(\mv x,y)\in Q(D)} y^2.
\end{eqnarray}
Then,
\vspace{-.6cm}
\begin{align}
   J(\vec\theta) &= \frac 1 2 g(\vec\theta)^\top \vec\Sigma g(\vec\theta)
   - \inner{g(\vec\theta), \mv c} + \frac{s_Y}{2}
   +\frac \lambda 2 \norm{\vec\theta}^2\label{eqn:point:eval}\\
   \grad J(\vec\theta) &= \pd{g(\vec\theta)^\top}{\vec\theta}\vec\Sigma
   g(\vec\theta)-\pd{g(\vec\theta)^\top}{\vec\theta} \mv c+\lambda\vec\theta\label{eqn:gradient}
\end{align}

The quantity $\pd{g(\vec\theta)^\top}{\vec\theta}$ is a $p \times m$ matrix,
and $\vec\Sigma$ is an $m \times m$ matrix of {\em sparse tensors}. 
Statistically, $\vec\Sigma$ is related to the covariance matrix, 
$\mv c$ to the correlation between the response and the regressors,
and $s_Y$ to the empirical second moment of the regressand. 
When all input features are continuous, each component function $h_j(\mv x)$ is a
monomial giving a scalar value. In real-world workloads there 
is always a mix of categorical and continuous features, where
the monomials $h_j(\mv x)$ become {\em tensor products}, and so
the quantities $h_i(\mv x) \otimes h_j(\mv x)$ are also tensor
products, represented by relational queries with group-by. 
The group-by variables for the aggregate query computing the sum of tensor products $\sum_{(\mv x,y)\in Q(D)} h_i(\mv x)\otimes h_j(\mv
   x)$ in $\vec\sigma_{ij}$ are
precisely the categorical variables occurring in the monomials defining the
component functions $h_i$ and $h_j$.
For example,
if $h_i(\mv x) = x_A$ and $h_j(\mv x) = x_B$, where $A$ and $B$ are
      continuous features, then there is no group-by variable and the above sum of tensor products 
      is expressed as follows in SQL:
\begin{verbatim}
      SELECT sum(A*B) from Q;
\end{verbatim}
On the other hand, if both $A$ and $B$ are
      categorical variables, then the SQL encoding has two group-by variables $A$ and
      $B$:
\begin{verbatim}
      SELECT A, B, count(*) FROM Q GROUP BY A, B;
\end{verbatim}

These aggregates exploit the sparsity of the representation of $\vec\sigma_{ij}$ over categorical features to achieve succinct representation: The group-by clause ensures that only combinations of categories for the query variables $A$ and $B$ that exist in the training dataset are considered. 
The aggregates $\vec c$ and $s_Y$ are treated similarly.

The above rewriting allows us to compute the data-dependent quantities $\vec\Sigma$, $\mv c$, and $s_Y$  in the loss function $J$ and its gradient $\grad J$ once for all iterations of AC/DC. They can be computed efficiently  
{\em inside the database} as aggregates over the query $Q$. Aggregates with different group-by clauses may require different evaluation strategies to attain the best known complexity~\cite{fdb,faq}. AC/DC settles instead for one strategy for all aggregates, cf.\@ Section~\ref{sec:evaluation}. 
This has two benefits. First, the underlying join is only computed once for all aggregates. Second, the computation of the aggregates can be shared massively. These benefits may easily dwarf the gain of using specialised evaluation strategies for individual aggregates in case of very many aggregates (hundreds of millions) and large databases.

{\noindent\bf Reparameterization under Functional Dependencies (FDs).}
AC/DC exploits the FDs among variables in the feature
extraction query $Q$ to reduce the dimensionality of the optimization problem by
eliminating functionally determined variables and re-parameterizing the model.
We thus only compute the quantities $\vec\Sigma$, $\mv c$, and $s_Y$ on the
subset of the features that are not functionally determined and solve the
lower-dimensional optimization problem. The effect on the loss function and its gradient is immediate and uniform across all optimization problems in our class: We have less terms to compute since the functionally determined variables are dropped. The effect on the non-linear penalty term $\Omega$ is however non-trivial and depends on the model at hand~\cite{NNOS2018}.

We next explain the reparameterization
of the ridge linear regression from Example~\ref{ex:lr} under the FD
$\textsf{city}\rightarrow \textsf{country}$~\cite{NNOS2018}. 
The categorical features
$\mv x_{\textsf{city}}$ and $\mv x_{\textsf{country}}$ are represented by
indicator vectors and the latter can be recovered from the former using the
mapping between values for \textsf{city} and \textsf{country} in the input
database. We can extract this map $R(\mathsf{country},\mathsf{city})$ that is a
sparse representation of a matrix $\mv R$ for which
$\mv x_{\mathsf{city}}= \mv R\mv x_{\mathsf{country}}$. The model
becomes:
\begin{align*}
    \inner{\vec\theta,\mv x}
   = \sum_{j\notin \{{\mathsf{city}},{\mathsf{country}}\}}\inner{\vec\theta_j,\mv x_j}+
   \inner{\underbrace{\vec\theta_{\mathsf{city}} + \mv R^\top
   \vec\theta_{\mathsf{country}}}_{\vec\gamma_{\mathsf{city}}},\mv x_{\mathsf{city}}}.
\end{align*}
The parameters $\vec\theta_{\textsf{city}}$ and $\vec\theta_{\textsf{country}}$
are replaced by new parameters $\vec\gamma_{\textsf{city}}$ for
the categorical features $\mv x_{\textsf{city}}$. This new form can be used in
the loss function~\eqref{eqn:general:form:of:J}. We can further optimize
out $\vec\theta_{\textsf{country}}$ from the non-linear penalty term by setting
the partial derivative of $J$ with respect to $\vec\theta_{\textsf{country}}$ to
0. Then, the new penalty term becomes:
\begin{align*}
\norm{\vec\theta}_2^2 
&= \sum_{j\neq \{{\mathsf{city}},{\mathsf{country}}\}} \!\!\!\!\!\norm{\vec\theta_j}_2^2 + \norm{\vec\gamma_{\textsf{city}} - \mv  R^\top \vec\theta_{\textsf{country}}}_2^2 + \norm{\vec\theta_{\textsf{country}}}_2^2 \\
&= \sum_{j\neq  \{{\mathsf{city}},{\mathsf{country}}\}} \!\!\!\!\!\norm{\vec\theta_j}_2^2 + \inner{(\mv I_{\textsf{city}} + \mv R^\top \mv R)^{-1}\vec\gamma_{\textsf{city}},\vec \gamma_{\textsf{city}}},
\end{align*}
where $\mv I_{\textsf{city}}$ is the identity matrix in the order of the active domain size
of \textsf{city}. A similar formulation of the penalty term holds for polynomial regression models. 

AC/DC thus exploits FDs to compute fewer aggregates at
the cost of a more complex regularizer term that requires 
matrix inversion and multiplication. While these matrix operations are expressible as
database queries and can be evaluated similarly to the other
aggregates~\cite{NNOS2018}, AC/DC uses instead the Eigen library for linear
algebra to achieve significant speedups for model reparameterization over the
strawman approach that does not exploit FDs.

\begin{figure*}[t]
  \begin{center}
    \begin{tabular}{|l|}\hline {\bf aggregates} (variable order $\Delta$, varMap,
      relation ranges $R_1[x_1,y_1],\ldots,R_d[x_d,y_d]$)\\\hline
      $A = root(\Delta); \TAB \text{context} = \pi_{dep(A)} (\text{varMap});
      \TAB \text{reset}(\text{aggregates}_A); \TAB \#\aggs = \left| \aggs_A \right|$;\\ \\
      $\IF\STAB (dep(A) \neq anc(A)) \TAB \{ 
      \TAB \text{aggregates}_A = \text{cache}_A[\text{context}]; \TAB \IF\STAB
      (\text{aggregates}_A[0]  \neq \emptyset) \STAB\RETURN; \STAB \}$\\ \\
      $\FOREACH\STAB i\in [d]\STAB\DO\STAB R_i[x'_i,y'_i] = R_i[x_i,y_i]$;\\
      $\FOREACH\STAB a\in \bigcap_{i\in[d] \mbox{ such that } A\in vars(R_i)} \pi_A (R_i[x_i,y_i])
      \STAB\DO\STAB \{$\\
      $\TAB\FOREACH\STAB i\in [d] \mbox{ such that } A\in vars(R_i)\STAB\DO\STAB\text{find range } R_i[x'_i,y'_i]
      \subseteq R_i[x_i,y_i] \mbox{ such that } \pi_A (R_i[x'_i,y'_i]) =
      \{(A:a)\};$\\
      $\TAB\MATCH\ (A):$ \\
      {\color{red}$\TAB\TAB$
      \begin{tabular}{|ll|}\hline
        {\color{black}$\textbf{continuous feature } : \STAB$}
        & {\color{black}$\lambda_A = [\{() \mapsto 1\}, \{() \mapsto a^1\} , 
          \ldots,\{()\mapsto a^{2 \cdot degree}\} \}];$}\\
        {\color{black}$\textbf{categorical feature } : \STAB$}
        & {\color{black}$\lambda_A = [\{() \mapsto 1\}, \{a \mapsto 1\}];$}\\
        {\color{black}$\textbf{no feature } : \STAB$} 
        & {\color{black}$\lambda_A = [\{() \mapsto 1\}];$}\\\hline
      \end{tabular}
      } \\
      $\TAB\MATCH\ (\Delta):$\\
      $\TAB\TAB \textbf{leaf node } A: $\\
      $\TAB\TAB\TAB$\rfbox{$\FOREACH\STAB l \in \left[\#\aggs\right]\STAB\DO\STAB\{$
      $\TAB [i_0] = \mathcal{R}_A[l]; \TAB \aggs_A[l] \pluseq \lambda_A[i_0]; \TAB\}$}\\
      $\TAB\TAB \textbf{inner node } A(\Delta_1, \ldots, \Delta_k): $\\
      $\TAB\TAB\TAB\FOREACH\STAB j\in [k]\STAB\DO\STAB 
      \text{\bf aggregates}(\Delta_j,\text{varMap}\times\{(A:a)\},
      \text{ranges } R_1[x'_1,y'_1], \ldots, R_d[x'_d,y'_d]);$\\
      {\color{red}$\TAB\TAB\TAB $
      \begin{tabular}{|l|}\hline
      {\color{black}
      $\IF\STAB (\forall j \in [k] : \aggs_{root(\Delta_j)}[0] \neq \emptyset)
      \STAB$}\\
        {\color{black}
        $\TAB\FOREACH\STAB l \in \left[\#\aggs\right]\;\DO\;\{\; [i_0, i_1, \ldots, i_k] = \mathcal{R}_A[l];\TAB \aggs_A[l] \pluseq 
        \lambda_A[i_0]\otimes\bigotimes_{j \in [k]}\aggs_{root(\Delta_j)}[i_j];\; \}$}\\\hline 
      \end{tabular}
      } \\
      $\}$\\ 
      $\IF\STAB (dep(A) \neq anc(A))\TAB \text{cache}_A[\text{context}] = 
      \aggs_A;$\\\hline
    \end{tabular}
  \end{center}
  \caption{Algorithm for computing aggregates
    $\aggs_A$. Each aggregate
    is a map from tuples over its group-by variables to scalars.
    The parameters of the initial call are the variable order
    $\Delta$ of the feature extraction query, an empty map from variables to values, and the full range of tuples for each
    relation $R_1,\ldots,R_d$ in the input database.
  } \label{fig:aggcomp}
\end{figure*}

\section{Aggregate Computation in AC/DC}
\label{sec:evaluation}

An immediate approach to computing the aggregates in $\mv \Sigma$, $\vec c$, and
$s_Y$ for a given optimization problem is 
to first materialize the result of the feature
extraction query $Q$ using an efficient query engine, e.g., a worst-case optimal
join algorithm, and then compute the aggregates in one pass over the query result.  
This approach, however, is suboptimal, since the listing
representation of the query result is highly redundant and not necessary for the
computation of the aggregates.  AC/DC avoids this redundancy by factorizing the
computation of aggregates over joins, as detailed in
Section~\ref{sec:factorized}.  In a nutshell, this factorized approach unifies
three powerful ideas: worst-case optimality for join processing, query plans
defined by fractional hypertree decompositions of join queries, and an
optimization that partially pushes aggregates past joins. AC/DC further exploits
similarities across the aggregates to massively share their computation, as
detailed in Section~\ref{sec:shared}.

\subsection{Factorized Computation of Aggregates} 
\label{sec:factorized}

Factorized aggregate computation relies on a variable
order $\Delta$ for the query $Q$ to avoid redundant computation. 
In this paper, we assume that we are given a variable order. Prior work discusses this query optimization problem~\cite{sigrec16,fdb,faq}.

{\bf Variable Orders.} State-of-the-art query evaluation uses relation-at-a-time query plans. 
We use variable-at-a-time query plans, which we call variable orders. These are partial orders on the variables in the query, capture the join dependencies in the
query, and dictate the order in which we solve each join variable. 
For each variable, we join all relations with that variable. 
Our choice is motivated by the
complexity of join evaluation: Relation-at-a-time query plans are
provably suboptimal, whereas variable-at-a-time query plans can be chosen to be optimal~\cite{skew}.

Given a join query $Q$, a variable $X$ \emph{depends} on a variable $Y$ if both
are in the schema of a relation in $Q$.

\begin{definition}[adapted from~\cite{OlZa15}]
  A variable order $\Delta$ for a join query $Q$ is a pair $(F, dep)$, where $F$
  is a rooted forest with one node per variable in $Q$, and $dep$ is a function
  mapping each variable $X$ to a set of variables in $F$. 
  It satisfies the following constraints:
  \begin{itemize}
  \item For each relation in $Q$, its variables lie along the same root-to-leaf
    path in $F$.
  \item For each variable $X$, $dep(X)$ is the subset of its ancestors in $F$ on which
    the variables in the subtree rooted at $X$ depend.
  \end{itemize}
\end{definition}

Without loss of generality, we use variables orders that are trees instead of
forests. We can convert a forest into a tree by adding to each relation the same dummy join variable that takes a single value.
For a variable $X$ in the variable order $\Delta$, $anc(X)$ is the set of all
ancestor variables of $X$ in $\Delta$.  The set of variables in $\Delta$ (schema
of a relation $R$) is denoted by $vars(\Delta)$ ($vars(R)$ respectively) and the
variable at the root of $\Delta$ is denoted by $root(\Delta)$.

\begin{example} 
  Figure~\ref{fig:varorder} shows a variable order for the natural join of
  relations $R(A,B,C)$, $T(B,D)$, and $S(A,E)$. 
  Then, $anc(D)=\{A,B\}$ and $dep(D)=\{B\}$, i.e., $D$ 
  has ancestors $A$ and $B$, yet it only depends on $B$. 
  Given $B$, the variables $C$ and $D$ are independent of each
  other. For queries with group-by variables, we choose a variable order
  where these variables sit above the other variables~\cite{fdb}.
\end{example}

Figure~\ref{fig:aggcomp} presents the AC/DC algorithm for factorized computation
of SQL aggregates over the feature extraction query $Q$. The 
backbone of the algorithm without the code in boxes explores the 
factorized join of the input relations $R_1,\ldots,R_d$ over a variable order $\Delta$ of $Q$. As it traverses $\Delta$ in depth-first preorder, it assigns values to the query variables. The assignments are kept in varMap and used to compute aggregates by the code in the boxes.

The relations are sorted following a depth-first
pre-order traversal of $\Delta$. Each call takes a range $[x_i,y_i]$ of tuples in
each relation $R_i$. Initially, these ranges span the entire relations. 
Once the root variable $A$ in $\Delta$ is assigned a value $a$ from the intersection of possible $A$-values from the input relations, these ranges are narrowed down to those tuples with value $a$ for $A$.

To compute an aggregate over the variable order $\Delta$ rooted at $A$, we first 
initialize the aggregate to zeros. This is needed since the aggregates might have been used earlier for different assignments of ancestor variables in $\Delta$.
We next check whether we previously computed the
aggregate for the same assignments of variables in $dep(A)$, denoted by $\text{context}$, and cached it in a map $\text{cache}_A$. 
Caching is useful when $dep(A)$ is strictly contained in $anc(A)$, 
since this means that the aggregate computed at $A$ does not need to be
recomputed for distinct assignments of variables in $anc(A)\setminus dep(A)$. 
In this case, we probe the cache using as key the assignments in varMap of the $dep(A)$ variables: $\text{cache}_A[\text{context}]$.
If we have already computed the aggregates over that assignment for $dep(A)$, 
then we can just reuse the previously computed aggregates and avoid recomputation. 

If $A$ is a group-by variable, then we compute a map from each $A$-value $a$ to a function of $a$ and aggregates computed at children of $A$, if any. If $A$ is not a group-by variable, then we compute a map from the empty value $()$ to such a function; in this latter case, we could have just computed the aggregate instead of the map though we use the map for uniformity. In case there are group-by variables under $A$, the computation at $A$ returns maps whose keys are tuples over all these group-by variables 
in $vars(\Delta)$.

\begin{example} \label{ex:aggpushdown} 
  Consider the query $Q$ with the variable order $\Delta$ in Figure~\ref{fig:varorder}. 
  We first compute the assignments for $A$ as
  $Q_A = \pi_A R \bowtie \pi_A T$. For each assignment $a \in Q_A$, we
  then find assignments for variables under $A$ within the narrow ranges
  of tuples that contain $a$. 
  The assignments for $B$ in the context of $a$ are given by 
  $Q^{a}_B = \pi_B(\sigma_{A = a}R) \bowtie \pi_BS$. For each
  $b \in Q^{a}_B$, the assignments for $C$ and $D$ are given by
  $Q^{a,b}_C = \pi_C(\sigma_{A = a \land B = b}R)$ and
  $Q^{b}_D = \pi_D(\sigma_{B = b}S)$. Since $D$ depends on $B$ and not on $A$, the 
  assignments for $D$ under a given $b$ are repeated for every occurrence of $b$
  with assignments for $A$. 
  The assignments for $E$ given $a\in Q_A$ are computed as
  $Q^{a}_E = \pi_E(\sigma_{A = a}T)$.

  Consider the aggregate $\COUNT(Q)$. 
  The count at each variable $X$ is computed
  as the sum over all value assignments of $X$ of the product of the counts 
  at the children of $X$ in $\Delta$; if $X$ is a leaf in $\Delta$, 
  the product at children is considered 1.
  For our variable order,
  this computation is captured by the following factorized expression:
  \begin{align}
    \COUNT =
    \sum_{a \in Q_A} \!\! 1 \times \left(\sum_{b \in Q^{a}_B} \!\! 1 \times
    \left(\sum_{c \in Q^{a,b}_C} \!\!\!1 \times V_D(b) \right)\right)
    \times \sum_{e \in Q^{a}_E} \!\! 1
    \label{eq:count}
  \end{align}
  where $V_D(b) = \sum_{d \in Q^{b}_D}\!\! 1$ is cached the first time 
  we encounter the assignment $b$ for $B$ and reused for all subsequent
  occurrences of this assignment under assignments for $A$.
  
  Summing all $X$-values in the result of $Q$ for a variable $X$ is done similarly, with the difference that at the variable $X$ in $\Delta$ we compute the sum of the values of $X$ weighted by the product of the counts of their children.
  For instance, the aggregate $\SUM(C*E)$ is computed over our variable order 
  by the following factorized expression:
  \begin{align}
    \SUM(C \cdot E) =
    \sum_{a \in Q_A} \!\! 1 \times \left(\sum_{b \in Q^{a}_B} \!\! 1 \times
    \left(\sum_{c \in Q^{a,b}_C} \!\!\!c \times V_D(b) \right)\right)
    \times \sum_{e \in Q^{a}_E} \!\! e
    \label{eq:sum_ce}
  \end{align}

  To compute the aggregate $\SUM(C*E) \texttt{ GROUP BY } A$, we compute $\SUM(C*E)$ for each assignment for $A$ instead of marginalizing away $A$. The result is a map from $A$-values to values of $\SUM(C*E)$.
  

\end{example}

A good variable order may include
variables that are not explicitly used in the optimization problem. 
This is the case of join variables whose presence in the variable order 
ensures a good factorization. For instance, if we remove the variable $B$ from
the variable order in Figure~\ref{fig:varorder}, 
the variables $C,D$ are no longer independent and we cannot factorize the
computation over $C$ and $D$. AC/DC exploits the conditional 
independence enabled by $B$, but computes no aggregate over $B$ if this is not
required in the problem.

\begin{figure*} 
  \begin{tikzpicture}[array/.style={rectangle split,rectangle split
      horizontal, rectangle split parts=#1, draw, anchor=center}]
   
    \tikzstyle{rarray}= [
    minimum height=2em,
    minimum width=2.5em, draw
    ]
    
    \node at (-4, 0) (A) {$A$};
    \node at (-5, -4) (D1) {$\Delta_1$} edge[-] (A);
    \node at (-3, -4) (D2) {$\Delta_k$} edge[-] (A);
    \node at (-4, -4) {$\cdots$};
    
    \node [rarray, color=red ,draw]at (10,-1) (prod){
      {\large \color{black}$\alpha \pluseq \alpha_0 \otimes
        \bigotimes_{j \in [k]} \alpha_{j}$}};

    \begin{scope}[start chain=1 going right,node distance=-0.15mm]
      \node [rarray,on chain=1,draw=none] at (4.41,0) {$\aggs_A = \; \cdots \;$};
      \node [rarray,dashed, on chain=1,draw] {};
      \node [rarray, on chain=1,draw] (alpha){$\alpha$};
      \node [rarray,dashed, on chain=1,draw] {};
      \node [rarray, on chain=1,draw=none] {$\cdots$};
    \end{scope}

    \begin{scope}[start chain=1 going right,node distance=-0.15mm]
      \node [rarray,on chain=1,draw=none] at (5,-2) {}; 
      \node [rarray, on chain=1,draw] (i0){$i_0$};
      \node [rarray, on chain=1,draw] (i1){$i_1$};
      \node [rarray, on chain=1,draw] (ic){$\cdots$};
      \node [rarray, on chain=1,draw] (ik){$i_k$};
    \end{scope}

    \begin{scope}[start chain=1 going right,node distance=-0.15mm]
      \node [rarray, on chain=1,draw=none]  at (-2,-0) (lambdaA) {$\lambda_A = \; \cdots \;$};
      \node [rarray, on chain=1,dashed,draw] (a1) {};
      \node [rarray, on chain=1,draw] (a2) {$\alpha_0$};
      \node [rarray, on chain=1,dashed,draw] (a3){};
      \node [rarray, on chain=1,draw=none] {$\cdots$};
    \end{scope}
       
    \begin{scope}[start chain=1 going right,node distance=-0.15mm]
      \node [rarray,on chain=1,draw=none,anchor=east] at (-1.5,-4) {};
      \node [rarray, on chain=1,draw=none] (delta1){$\aggs_{root(\Delta_1)} = \; \cdots \;$};
      \node [rarray, on chain=1,dashed,draw] (i11){};
      \node [rarray, on chain=1,draw] (i12){$\alpha_{1}$};
      \node [rarray, on chain=1,dashed,draw] (i13){};
      \node [rarray, on chain=1,draw=none] {$\cdots$};
    \end{scope}

    \begin{scope}[start chain=1 going right,node distance=-0.15mm]
      \node [rarray, on chain=1,draw=none]at (5.5,-4) {};
      \node [rarray, on chain=1,draw=none] (delta2){$\aggs_{root(\Delta_k)} = \; \cdots$};
      \node [rarray, on chain=1,dashed,draw] (ik1){};
      \node [rarray, on chain=1,draw] (ik2){$\alpha_{k}$};
      \node [rarray, on chain=1,dashed,draw] (ik3){};
      \node [rarray, on chain=1,draw=none] {$\cdots$};
    \end{scope}

    \node[color=black!80, anchor=south, yshift=-1] at (a1.north) {\scriptsize $i_0 - 1$};
    \node[color=black!80, anchor=south, yshift=-1] at (a2.north) {\scriptsize $i_0$};
    \node[color=black!80, anchor=south, yshift=-1] at (a3.north) {\scriptsize $i_0 + 1$};

    \node[color=black!80, anchor=south, yshift=-1] at (i11.north) {\scriptsize $i_1 - 1$};
    \node[color=black!80, anchor=south, yshift=-1] at (i12.north) (ii1) {\scriptsize $i_1$};
    \node[color=black!80, anchor=south, yshift=-1] at (i13.north)  {\scriptsize $i_1 + 1$};

    \node[color=black!80, anchor=south, yshift=-1] at (ik1.north) {\scriptsize $i_k - 1$};
    \node[color=black!80, anchor=south, yshift=-1] at (ik2.north) (iik) {\scriptsize $i_k$};
    \node[color=black!80, anchor=south, yshift=-1] at (ik3.north)  {\scriptsize $i_k + 1$};

    \draw[->] (alpha.south) -- (ic.north west);

    \draw[->] (i0.south) -- ($ (i0.south) + (0,-0.3) $) --  ($ (a2.south) + (0,-2.3) $)
    -- (a2.south);
    \draw[->] (i1.south) -- ($ (i1.south) + (0,-0.46) $) --  ($ (ii1.north) + (0,0.5) $)
    -- (ii1.north);
    \draw[->] (ik.south) -- ($ (ik.south) + (0,-0.46) $) --  ($ (iik.north) + (0,0.5) $)
    -- (iik.north);
   
  \end{tikzpicture}%
  \caption{Index structure provided by the aggregate register for a particular
    aggregate $\alpha$ that is computed over the variable order
    $\Delta = A(\Delta_1, \ldots, \Delta_k)$. The computation of
    $\alpha$ is expressed as the sum of the Cartesian products of its aggregate components provided by the indices $i_0,\ldots,i_k$.} 
    \label{fig:genaggreg} \vspace*{-1em}
\end{figure*}
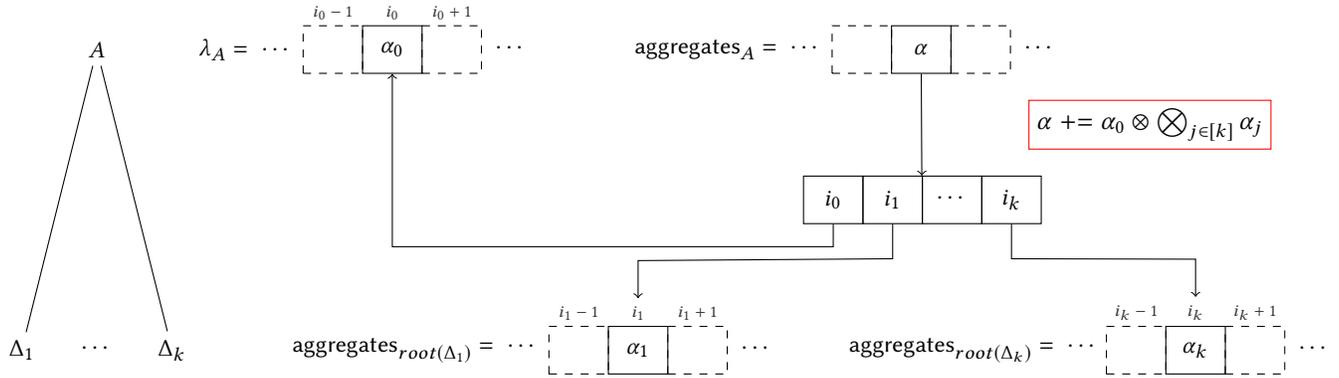

\subsection{Shared Computation of Aggregates}
\label{sec:shared}

Section~\ref{sec:factorized} explains how to factorize the computation of one
aggregate in $\vec\Sigma$, $\vec c$, and $s_Y$ over the join of database relations. In this section we show how to share the computation across these aggregates.

\begin{example}
  Let us consider the factorized expression of the sum aggregates $\SUM(C)$ and
  $\SUM(E)$ over $\Delta$:
  \begin{align}
    \SUM(C) &=
              \sum_{a \in Q_A} \!\! 1 \times \left(\sum_{b \in Q^{a}_B} \!\! 1
              \times \left(\sum_{c \in Q^{a,b}_C} \!\!\!c \times V_D(b)\right)\right) \times \sum_{e \in Q^{a}_E} \!\! 1 \label{eq:sum_c}\\
    \SUM(E) &=
              \sum_{a \in Q_A} \!\! 1 \times \left(\sum_{b \in Q^{a}_B} \!\! 1 \times
              \left(\sum_{c \in Q^{a,b}_C} \!\!\!1 \times V_D(b)\right)\right)
              \times \sum_{e \in Q^{a}_E} \!\! e \label{eq:sum_e}
  \end{align}
  We can share computation across the expressions~\eqref{eq:count} to \eqref{eq:sum_e} since they are similar.
  For instance, given an assignment $b$ for $B$, all these aggregates need $V_D(b)$. Similarly, for a given assignment $a$ for $A$, the aggregates \eqref{eq:sum_ce} and \eqref{eq:sum_e} can share the computation of
  the sum aggregate over $Q_E^{a}$. For assignments $a \in Q_A$ and
  $b \in Q^{a}_B$, \eqref{eq:sum_ce} and \eqref{eq:sum_c} can share the
  computation of the sum aggregate over $Q_C^{a,b}$.
\end{example}

AC/DC computes all aggregates together over a single variable order. It then shares as much computation as possible and significantly improves the data locality of the aggregate computation. AC/DC thus decidedly sacrifices the goal of achieving the lowest known complexity for individual aggregates for the sake of sharing as much computation as possible across these aggregates. 

{\noindent\bf Aggregate Decomposition and Registration.}  For a model of degree
$degree$ and a set of variables $\{A_l\}_{l\in[n]}$, we have aggregates of the
form $\SUM(\prod_{l\in[n]}A_l^{d_l})$, possibly with a group-by clause, such
that $0\leq \sum_{l\in[n]} d_l \leq 2\cdot degree$, $d_l \geq 0$, and all
categorical variables are turned into group-by variables. The reason for
$2\cdot degree$ is due to the $\vec\Sigma$ matrix used to compute the
gradient of the loss function~\eqref{eqn:gradient}, which pairs
any two features of degree up to $degree$.  Each aggregate is thus defined
uniquely by a monomial $\prod_{l\in[n]}A_l^{d_l}$; we may discard the variables
with exponent 0. For instance, the monomial for $\SUM(C*E)$ is \texttt{CE} while for $\SUM(C*E) \text{ GROUP BY } A$ is \texttt{{\bf A}CE}.

Aggregates can be decomposed into shareable components. Consider a variable order $\Delta = A(\Delta_1, \ldots, \Delta_k)$, with root $A$ and subtrees $\Delta_1$ to $\Delta_k$.
We can decompose any aggregate $\alpha$ to be
computed over $\Delta$ into $k+1$
aggregates such that aggregate $0$ is for $A$ and aggregate $j \in [k]$ is for $root(\Delta_j)$. Then $\alpha$ is computed as the product of its $k+1$ components. Each of these aggregates is defined by the projection of the monomial of $\alpha$ onto $A$ or $vars(\Delta_j)$. 
The aggregate $j$ is then pushed down the variable order and computed over the subtree $\Delta_j$. If the projection of the
monomial is empty, then the aggregate to be pushed down is $\SUM(1)$, which
computes the size of the join defined by $\Delta_j$. If
several aggregates push the same aggregate to the subtree $\Delta_j$, this
is computed only once for all of them. 

The decomposed aggregates form a hierarchy whose structure is that of the underlying variable order $\Delta$. The aggregates at a variable $X$ are denoted by $\aggs_X$. All aggregates are to be computed at the root of $\Delta$, then fewer are computed at each of its children and so on. This structure is the same regardless of the input data and can be constructed before data processing.
We therefore construct at compile time for each variable $X$ in $\Delta$ an aggregate register $\mathcal{R}_X$ that is an array of 
all aggregates to be computed  over the subtree of $\Delta$ rooted at $X$.
This register is used as an index structure to facilitate the computation of the
actual aggregates. More precisely, an entry for an aggregate $\alpha$ in the register of $X$ is labeled by the monomial of $\alpha$ and holds an array of indices of the components of $\alpha$ located in the registers at the children of $X$ in $\Delta$ and in the local register $\Lambda_X$ of $X$. 
Figure~\ref{fig:genaggreg} depicts this construction.

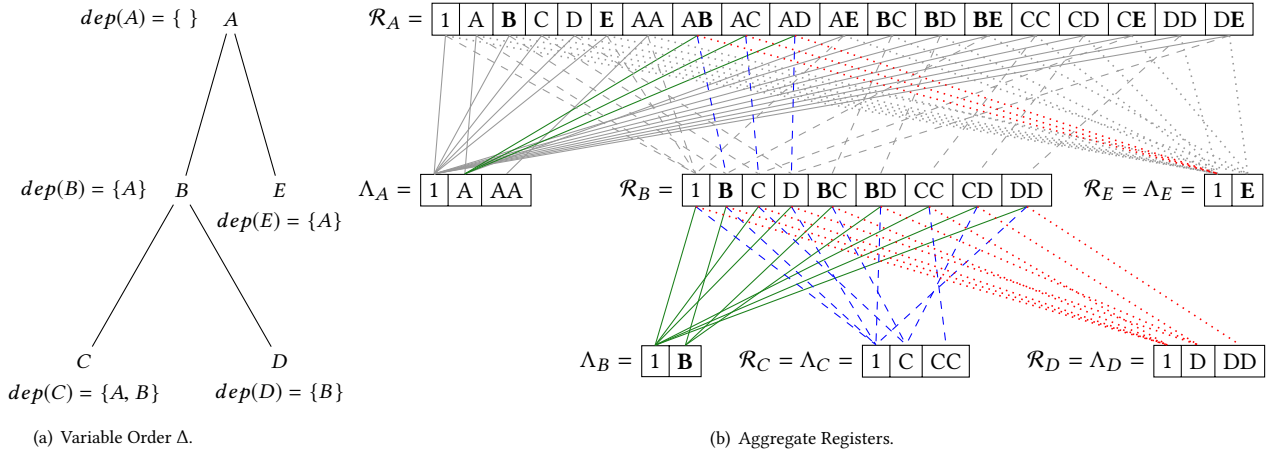
\begin{figure*} 
  \subfigure[\label{fig:varorder}Variable Order $\Delta$.]{
    \begin{minipage}{3cm}
      \begin{small}
        \begin{tikzpicture}[xscale=0.65, yscale=0.65]

          \node at (-3, -1) (A) {$A$};
          \node at (-4, -4.5) (B) {$B$} edge[-] (A);
          \node at (-2, -8) (D) {$D$} edge[-] (B);
          \node at (-6, -8) (C) {$C$} edge[-] (B);
          \node at (-2, -4.5) (E) {$E$} edge[-] (A);

          \begin{small}
            \node[anchor = north] at (C.south) {$dep(C) = \{A,B\}$};
            \node at (-6, -4.5) {$dep(B) = \{A\}$};
            \node[anchor = north] at (D.south) {$dep(D) = \{B\}$};
            \node[anchor = north] at (E.south) {$dep(E) = \{A\}$};
            \node[anchor = east] at (-3.5, -1) {$dep(A) = \{\;\}$};
          \end{small}

          \node at (5,-9.2) (xxx) {~};

        \end{tikzpicture}%
      \end{small}
    \end{minipage} }\hspace{1.5cm}%
  \subfigure[\label{fig:aggreg}Aggregate Registers.]{
    \begin{minipage}{12cm}
      \begin{tikzpicture}[array/.style={rectangle split,rectangle split
          horizontal, rectangle split parts=#1,draw, anchor=center},xscale=0.65,
        yscale=0.65]

        \node[array=19] at (-2.5, 0) (aggA) {
          \nodepart{one} 1
          \nodepart{two} A
          \nodepart{three} {\bf B}
          \nodepart{four} C
          \nodepart{five} D
          \nodepart{six} {\bf E}
          \nodepart{seven} AA
          \nodepart{eight} A{\bf B}
          \nodepart{nine} AC
          \nodepart{ten} AD
          \nodepart{eleven} A{\bf E}
          \nodepart{twelve} {\bf B}C
          \nodepart{thirteen} {\bf B}D
          \nodepart{fourteen} {\bf B}{\bf E}
          \nodepart{fifteen} CC
          \nodepart{sixteen} CD
          \nodepart{seventeen} C{\bf E}
          \nodepart{eighteen} DD
          \nodepart{nineteen} D{\bf E}
        };

        \node[anchor=east] at (aggA.one west) (talphaA) {$\mathcal{R}_A = $};

        \node[array=3] at (-10, -3.5) (lambdaA) {
          \nodepart{one} 1
          \nodepart{two} A
          \nodepart{three} AA
        };

        \node[anchor=east] at (lambdaA.one west) (tlambaA) {$\Lambda_A = $};

        \node[array=9] at (-2, -3.5) (aggB) {
          \nodepart{one} 1
          \nodepart{two} {\bf B}
          \nodepart{three} C
          \nodepart{four} D
          \nodepart{five} {\bf B}C
          \nodepart{six} {\bf B}D
          \nodepart{seven} CC
          \nodepart{eight} CD
          \nodepart{nine} DD
        };

        \node[anchor=east] at (aggB.one west) (talphaA) {$\mathcal{R}_B = $};
        
        \node[array=2] at (5.5, -3.5) (aggE) { \nodepart{one} 1 \nodepart{two}
          {\bf E} };

        \node[anchor=east] at (aggE.one west) (talphaA) {$\mathcal{R}_E = \Lambda_E = $};

        \node[array=2] at (-6, -7) (lambdaB) { \nodepart{one} 1
          \nodepart{two} {\bf B} };

        \node[anchor=east] at (lambdaB.one west) (talphaA) {$\Lambda_B = $};

        \node[array=3] at (-1, -7) (aggC) { \nodepart{one} 1 \nodepart{two} C
          \nodepart{three} CC };
        
        \node[anchor=east] at (aggC.one west) (talphaA) {$\mathcal{R}_C = \Lambda_C = $};

        \node[array=3] at (5, -7) (aggD) { \nodepart{one} 1 \nodepart{two} D
          \nodepart{three} DD };

        \node[anchor=east] at (aggD.one west) (talphaA) {$\mathcal{R}_D = \Lambda_D = $};

        \draw[color=black!40] (aggA.one south) -- (lambdaA.one north);
        \draw[color=black!40] (aggA.two south) -- (lambdaA.two north);
        \draw[color=black!40] (aggA.three south) -- (lambdaA.one north);
        \draw[color=black!40] (aggA.four south) -- (lambdaA.one north);
        \draw[color=black!40] (aggA.five south) -- (lambdaA.one north);
        \draw[color=black!40] (aggA.six south) -- (lambdaA.one north);
        \draw[color=black!40] (aggA.seven south) -- (lambdaA.three north);
        \draw[color=black!40] (aggA.eleven south) -- (lambdaA.two north);
        \draw[color=black!40] (aggA.twelve south) -- (lambdaA.one north);
        \draw[color=black!40] (aggA.thirteen south) -- (lambdaA.one north);
        \draw[color=black!40] (aggA.fourteen south) -- (lambdaA.one north);
        \draw[color=black!40] (aggA.fifteen south) -- (lambdaA.one north);
        \draw[color=black!40] (aggA.sixteen south) -- (lambdaA.one north);
        \draw[color=black!40] (aggA.seventeen south) -- (lambdaA.one north);
        \draw[color=black!40] (aggA.eighteen south) -- (lambdaA.one north);
        \draw[color=black!40] (aggA.nineteen south) -- (lambdaA.one north);

        \draw[color=goodgreen] (aggA.eight south) -- (lambdaA.two north);
        \draw[color=goodgreen] (aggA.nine south) -- (lambdaA.two north);
        \draw[color=goodgreen] (aggA.ten south) -- (lambdaA.two north);

        \draw[color=black!40, dashed] (aggA.one south) -- (aggB.one north);
        \draw[color=black!40, dashed] (aggA.two south) -- (aggB.one north);
        \draw[color=black!40, dashed] (aggA.three south) -- (aggB.two north);
        \draw[color=black!40, dashed] (aggA.four south) -- (aggB.three north);
        \draw[color=black!40, dashed] (aggA.five south) -- (aggB.four north);
        \draw[color=black!40, dashed] (aggA.six south) -- (aggB.one north);
        \draw[color=black!40, dashed] (aggA.seven south) -- (aggB.one north);
        \draw[color=black!40, dashed] (aggA.eleven south) -- (aggB.one north);
        \draw[color=black!40, dashed] (aggA.twelve south) -- (aggB.five north);
        \draw[color=black!40, dashed] (aggA.thirteen south) -- (aggB.six north);
        \draw[color=black!40, dashed] (aggA.fourteen south) -- (aggB.two north);
        \draw[color=black!40, dashed] (aggA.fifteen south) -- (aggB.seven north);
        \draw[color=black!40, dashed] (aggA.sixteen south) -- (aggB.eight north);
        \draw[color=black!40, dashed] (aggA.seventeen south) -- (aggB.three north);
        \draw[color=black!40, dashed] (aggA.eighteen south) -- (aggB.nine north);
        \draw[color=black!40, dashed] (aggA.nineteen south) -- (aggB.four north);

        \draw[color=blue, dashed] (aggA.eight south) -- (aggB.two north);
        \draw[color=blue, dashed] (aggA.nine south) -- (aggB.three north);
        \draw[color=blue, dashed] (aggA.ten south) -- (aggB.four north);

        \draw[color=black!40,semithick,dotted] (aggA.one south) -- (aggE.one north);
        \draw[color=black!40,semithick,dotted] (aggA.two south) -- (aggE.one north);
        \draw[color=black!40,semithick,dotted] (aggA.three south) -- (aggE.one north);
        \draw[color=black!40,semithick,dotted] (aggA.four south) -- (aggE.one north);
        \draw[color=black!40,semithick,dotted] (aggA.five south) -- (aggE.one north);
        \draw[color=black!40,semithick,dotted] (aggA.six south) -- (aggE.two north);
        \draw[color=black!40,semithick,dotted] (aggA.seven south) -- (aggE.one north);
        \draw[color=black!40,semithick,dotted] (aggA.eleven south) -- (aggE.two north);
        \draw[color=black!40,semithick,dotted] (aggA.twelve south) -- (aggE.one north);
        \draw[color=black!40,semithick,dotted] (aggA.thirteen south) -- (aggE.one north);
        \draw[color=black!40,semithick,dotted] (aggA.fourteen south) -- (aggE.two north);
        \draw[color=black!40,semithick,dotted] (aggA.fifteen south) -- (aggE.one north);
        \draw[color=black!40,semithick,dotted] (aggA.sixteen south) -- (aggE.one north);
        \draw[color=black!40,semithick,dotted] (aggA.seventeen south) --
        (aggE.two north);
        \draw[color=black!40,semithick,dotted] (aggA.eighteen south) -- (aggE.one north);
        \draw[color=black!40,semithick,dotted] (aggA.nineteen south) -- (aggE.two north);

        \draw[color=red,semithick,dotted] (aggA.eight south) -- (aggE.one north);
        \draw[color=red,semithick,dotted] (aggA.nine south) -- (aggE.one north);
        \draw[color=red,semithick,dotted] (aggA.ten south) -- (aggE.one north);
        
        \draw[color = goodgreen] (aggB.one south) -- (lambdaB.one north);
        \draw[color = goodgreen] (aggB.two south) -- (lambdaB.two north);
        \draw[color = goodgreen] (aggB.three south) -- (lambdaB.one north);
        \draw[color = goodgreen] (aggB.four south) -- (lambdaB.one north);
        \draw[color = goodgreen] (aggB.five south) -- (lambdaB.two north);
        \draw[color = goodgreen] (aggB.six south) -- (lambdaB.two north);
        \draw[color = goodgreen] (aggB.seven south) -- (lambdaB.one north);
        \draw[color = goodgreen] (aggB.eight south) -- (lambdaB.one north);
        \draw[color = goodgreen] (aggB.nine south) -- (lambdaB.one north);

        \draw[color = blue, dashed] (aggB.one south) -- (aggC.one north);
        \draw[color = blue, dashed] (aggB.two south) -- (aggC.one north);
        \draw[color = blue, dashed] (aggB.three south) -- (aggC.two north);
        \draw[color = blue,dashed] (aggB.four south) -- (aggC.one north);
        \draw[color = blue, dashed] (aggB.five south) -- (aggC.two north);
        \draw[color = blue, dashed] (aggB.six south) -- (aggC.one north);
        \draw[color = blue, dashed] (aggB.seven south) -- (aggC.three north);
        \draw[color = blue, dashed] (aggB.eight south) -- (aggC.two north);
        \draw[color = blue, dashed] (aggB.nine south) -- (aggC.one north);
        
        \draw[color=red,semithick,dotted] (aggB.one south) -- (aggD.one north);
        \draw[color=red,semithick,dotted] (aggB.two south) -- (aggD.one north);
        \draw[color=red,semithick,dotted] (aggB.three south) -- (aggD.one north);
        \draw[color=red,semithick,dotted] (aggB.four south) -- (aggD.two north);
        \draw[color=red,semithick,dotted] (aggB.five south) -- (aggD.one north);
        \draw[color=red,semithick,dotted] (aggB.six south) -- (aggD.two north);
        \draw[color=red,semithick,dotted] (aggB.seven south) -- (aggD.one north);
        \draw[color=red,semithick,dotted] (aggB.eight south) -- (aggD.two north);
        \draw[color=red,semithick,dotted] (aggB.nine south) -- (aggD.three north);

        \node at (5,-8.2) (xxx) {~};
      \end{tikzpicture}%
    \end{minipage} }
  \caption{ (a) Variable order $\Delta$ for the natural join of the relations
    R(A,B,C), S(B,D), and
    T(A,E); (b) Aggregate registers for the aggregates needed to compute a linear regression model with degree 1 over $\Delta$. Categorical variables are shown in bold.
    }
  \vspace*{-1em}
  \label{fig:aggregateRegister}
\end{figure*}

The hierarchy of registers in Figure~\ref{fig:aggreg} forms an index structure that is used by AC/DC to compute the aggregates. This index structure is stored as one contiguous array in memory, where the entry for an aggregate $\alpha$ in the register comes with an auxiliary array with the indices of $\alpha$'s  aggregate components. The aggregates are ordered in the register so that we increase sequential access, and thus cache locality, when updating them.

\begin{example} \label{ex:aggreg} Let us compute a regression
  model of degree $1$ over a dataset defined by the join of
  the relations $R(A,B,C), S(B,D)$, and $T(A,E)$.  We assume that $B$ and $E$
  are categorical features, and all other variables are
  continuous. The quantities ($\vec \Sigma$,$\vec c$,$s_Y$) require the
  computation of the following aggregates: $\SUM(1)$, $\SUM(X)$ for each variable $X$, and $\SUM(X * Y)$ for each pair of variables $X$ and $Y$.
  
  Figure~\ref{fig:varorder} depicts a variable order $\Delta$ for the 
  natural join of three relations, and Figure~\ref{fig:aggreg} illustrates 
  the aggregate register that assigns a list of aggregates to each variable 
  in $\Delta$. 
  The aggregates are identified by their respective monomials (the names in the register entries). The categorical
  variables are shown in bold. 
  Since they are treated as group-by variables, we do not need aggregates whose
  monomials include categorical variables with exponents higher than 1. Any such
  aggregate is equivalent to the aggregate whose
  monomial includes the categorical variable with degree 1 only.

  The register $\mathcal{R}_A$ for the root $A$ of $\Delta$  
  has all aggregates needed to compute the model. 
  The register $\mathcal{R}_B$ has all aggregates from $\mathcal{R}_A$ 
  defined over the variables in the subtree of $\Delta$ rooted at $B$. 
  The variables $C$, $D$, and $E$ are leaf nodes in 
  $\Delta$, so the monomials for the aggregates in the registers $\mathcal{R}_C$, 
  $\mathcal{R}_D$, and $\mathcal{R}_E$ are the respective variables only.
  We use two additional registers $\Lambda_A$ and $\Lambda_B$, 
  which hold the aggregates corresponding to projections of the monomials 
  of the aggregates in $\mathcal{R}_A$, and respectively $\mathcal{R}_B$, onto 
  $A$, respectively $B$. For a leaf node $X$, the registers $\Lambda_X$ and $
  \mathcal{R}_X$ are the same.
  
  A path between two register entries in Figure~\ref{fig:aggreg} indicates 
  that the aggregate in the register above uses the result of the aggregate
  in the register below. For instance, each aggregate in $\mathcal{R}_B$ is
  computed by the product of one aggregate from $\Lambda_B$,
  $\mathcal{R}_C$, and $\mathcal{R}_D$. The fan-in of a register entry thus 
  denotes the amount of sharing of its aggregate: All aggregates from registers 
  above with incoming edges to this aggregate share its computation.
  For instance, the aggregates with monomials \texttt{AB}, \texttt{AC},
  and \texttt{AD} from $\mathcal{R}_A$ share the computation of the aggregate 
  with monomial \texttt{A} from $\Lambda_A$ as well as the count aggregate from 
  $\mathcal{R}_E$. 
  Their computation uses a sequential pass over the 
  register $\mathcal{R}_B$. This improves performance and access locality 
  as $\mathcal{R}_B$ can be stored in cache and accessed to compute all these aggregates. 
  \end{example}

{\noindent\bf Aggregate Computation.}  Once the aggregate registers are in place, we can ingest the input database and compute the aggregates over the join of the database relations following the factorized structure given by a variable order.
The algorithm in Figure~\ref{fig:aggcomp} does precisely this. Section~\ref{sec:factorized} explained the factorized computation of a single aggregate over the join. We explain here the case of several aggregates organized into the aggregate registers. This is stated by the pseudocode in the red boxes.

Each aggregate is uniformly stored as a map from tuples over their categorical variables to payloads that represent the sums over the projection of its monomial on all continuous variables. 
If the aggregate has no categorical variables, the key is the empty tuple.

For each possible $A$-value $a$, we first compute the array $\lambda_A$ that consists of the projections of the monomials of the aggregates onto $A$. If $A$ is categorical, then we only need to compute the 0 and 1 powers of $a$. If $A$ is continuous, we need to compute all powers of $A$ from 0 to $2\cdot degree$. If $A$ is not a feature used in the model, then we only compute a trivial count aggregate.

We update the value of each aggregate $\alpha$ using the index structure depicted in Figure~\ref{fig:genaggreg} as we traverse the variable order bottom up.  Assume we are at a variable $A$ in the variable order. In case $A$ is a leaf, the update is only a specific value in the local register $\lambda_A$.
In case the variable $A$ has children in the variable order, the aggregate is updated with the Cartesian product of all its component aggregates, i.e., one value from $\lambda_A$ and one aggregate for each child of $A$. The update value  can be expressed in SQL as follows. Assume the aggregate $\alpha$ has group-by variables $C$, which are partitioned across $A$ and its $k$ children. Assume also that $\alpha$'s components are $\alpha_0$ and $(\alpha_j)_{j\in[k]}$. Recall that all aggregates are maps, which we may represent as relations with columns for keys and one column $P$ for payload. Then, the update to $\alpha$ is:
\begin{align*}
  \texttt{SELECT } \mathcal{C}, (\alpha_0.P * \ldots
    * \alpha_k.P) \texttt{ AS } P
  \texttt{ FROM } \alpha_0, \ldots, \alpha_k;
\end{align*}

{\noindent\bf Further Considerations.} The auxiliary arrays that provide the precomputed indices of aggregate components within registers speed up the computation of the aggregates. Nevertheless, they still represent one extra level of indirection since each update to an aggregate would first need to fetch the indices and then use them to access the aggregate components in registers that may not be necessarily in the cache. We have been experimenting with an aggressive aggregate compilation approach that resolves all these indices at compile time and generates the specific code for each aggregate update. In experiments with linear regression, this compilation leads to a 4$\times$ performance improvements. However, the downside is that the AC/DC code gets much larger and the C++ compiler needs much more time to compile it. For higher-degree models, it can get into situations where the C++ compiler crashes. We are currently working on a hybrid approach that partially resolves the indices while maintaining a reasonable code size.

\section{The inner loop of BGD}
\label{sec:gradient}

As shown in Section~\ref{sec:theory}, the gradient descent solver repeatedly
computes $J(\vec\theta)$ and $\grad J(\vec\theta)$, which require matrix-vector
and vector-vector multiplications over the quantities
($\vec\Sigma,\vec c, s_Y$). We discuss here the computation that involves the
$\vec\Sigma$ matrix.

It is possible that several entries $\sigma_{ij} \in \vec\Sigma$ map to the same
aggregate query that is computed by AC/DC. Consider, for instance, the following
scalar entries in the feature mapping vector $h$: $h_i(\mv x) = x_a$,
$h_j(\mv x) = x_b\cdot x_c$, $h_k(\mv x) = x_b$, $h_l(\mv x) = x_a\cdot x_c$,
$h_m(\mv x) = x_c$, and $h_n(\mv x) = x_a\cdot x_b$. By definition of
$\vec\Sigma$, any pair-wise product of these entries in $h$ corresponds to one
entry in $\vec\Sigma$. The entries $\sigma_{ij}$, $\sigma_{lk}$, and
$\sigma_{mn}$ (as well as their symmetric counterparts) all map to the same
aggregate
$\SUM(A\cdot B\cdot C)=\sum_{(\mv x, y) \in Q(D)} x_a\cdot x_b \cdot x_c$.  To
avoid this redundancy, we use a sparse representation of $\vec\Sigma$, which
assigns to each distinct aggregate query a list of index pairs $(i,j)$ that
contains one pair for each entry $\sigma_{ij} \in \vec\Sigma$ that maps to this
query.

AC/DC operates directly over the sparse representation of $\vec\Sigma$. Consider
the matrix vector product $\mv p = \vec\Sigma g(\vec\theta)$, and let $A$ be the
root of the variable order $\Delta$. We compute $\mv p$ by iterating over all
aggregate maps $\alpha \in \aggs_{A}$, and for each index pair $(i,j)$ in $\vec\Sigma$ that is
assigned to $\alpha$, we add to the $i$'s entry in $\mv p$ with the product of
$\alpha$ and $j$'s entry in $g(\vec\theta)$. If $i \neq j$, we also add to $j$'s
entry in $\mv p$ with the product of $\alpha$ and $i$'s entry
in $g(\vec\theta)$.

{\bf Regularizer under FDs.}  Section~\ref{sec:theory} explains how to rewrite
the regularizer for a ridge linear regression model under the FD
$\mathsf{city} \rightarrow \mathsf{country}$. First, we need to construct the
relation $\mv R(\mathsf{country},\mathsf{city})$, and then compute the inverse
of the matrix $\mv D = (\mv I_{\mathsf{city}} + \mv R^\top \mv R)$. Note that each 
entry $(i,j)$ in $\mv R^\top \mv R$ is $1$, if two cities $\mathsf{city}_i$ and
$\mathsf{city}_j$ are in the same $\mathsf{country}$. Otherwise, the entry
$(i,j)$ is zero.

To facilitate the construction of the matrix $\mv D$, we construct $\mv R$ as a
map that groups the tuples of $\mv R$ by \textsf{country}. Thus, we construct a
mapping from \textsf{country} to the set of cities in this country that occur in
the dataset, which can be computed during the computation of the factorized
aggregates over the variable order.

We then use this representation of $\mv R$ to iterate over the payload set, and
for any two cities $\mathsf{city}_i, \mathsf{city}_j$ in the payload set, we
increment the corresponding index $(i,j)$ in $\mv D$ by one. We store $\mv D$
as a sparse matrix in the format used by the Eigen linear algebra library, 
and then use Eigen's Sparse Cholesky Decomposition to compute the inverse of
$\mv D$ and ultimately the solution for the regularizer.

\section{Experiments}
\label{sec:experiments}

We report on the performance of learning regression models and factorization
machines over a real dataset used in retail applications; 
cf.\@ the extended technical report~\cite{acdc} for
further experiments.

{\bf Systems.} We consider two variants of our system: The plain AC/DC and its
extension AC/DC+FD that exploits functional dependencies. We also report on five
competitors: F learns linear regression models and one-hot encodes the
categorical features~\cite{SOC16}; MADlib~\cite{MADlib:2012} 1.8 uses {\em ols}
to compute the closed-form solution of polynomial regression models (MADlib also
supports generalized linear models, but this is consistently slower than {\em
  ols} in our experiments and we do not report it here); R~\cite{R-project}
3.0.2 uses {\em lm} (linear model) based on QR-decomposition~\cite{Francis61};
libFM~\cite{libfm} 1.4.2 supports factorization machines; and
TensorFlow~\cite{tensorflow} 1.6 uses the LinearRegressor estimator with {\em
  ftrl} optimization~\cite{ftlr}, which is based on the conventional SGD optimization
algorithm.

The competitors come with strong limitations. MADlib inherits the limitation of
at most 1600 columns per relation from its PostgreSQL host. The MADlib one-hot
encoder transforms a categorical variable with $n$ distinct values into $n$
columns.  Therefore, the number of distinct values across all categorical
variables plus the number of continuous variables in the input data cannot
exceed 1600. R limits the number of values in their data frames to $2^{31}-1$.
There exist R packages, e.g., {\em ff}, which work around this limitation by
storing data structures on disk and mapping only chunks of data in main
memory. The biglm package can compute the regression model by processing one
ff-chunk at a time. Chunking the data, however, can lead to rank deficiencies
within chunks (feature interactions missing from chunks), which causes biglm to
fail. Biglm fails in all our experiments due to this limitation, and, thus, we
are unable to benchmark against it. LibFM requires as input a zero-suppressed
encoding of the join result. Computing this representation is an expensive
intermediary step between exporting the query result from the database system
and importing the data. To compute the model, we used its more stable MCMC
variant with a fixed number of runs (300); its SGD implementation requires a
fixed learning rate $\alpha$ and does not converge. AC/DC uses the adaptive
learning rate from Algorithm~\ref{algo:bgd} and runs until the parameters have
converged with high accuracy (for $\fama$, it uses 300 runs).

TensorFlow uses a user-defined iterator interface to load a batch of tuples from
the training dataset at a time.  This iterator defines a mapping from input
tuples to (potentially one-hot encoded) features and is called directly by the
learning algorithm. Learning over batches requires a random shuffling of the
input data, which in TensorFlow requires loading the entire dataset into
memory. This failed for our experiments and we therefore report its performance
without shuffling the input data.  We benchmark TensorFlow for $\lr$ only as it
does not provide functionality to create all pairwise interaction terms for
$\pr$ and $\fama$, third-party implementations of these models relied on python
packages that failed to load our datasets.  The optimal batch size for our
experiments is 100,000 tuples.  Smaller batch sizes require loading too many
batches, very large batches cannot fit into memory. Since TensorFlow requires a
fixed number of iterations, we report the times to do one epoch over the dataset
(i.e., computing 840 batches).  This means that the algorithm learned over each
input tuple once. In practice, it is often necessary to optimize with several epochs to
get a good model.

{\bf Experimental Setup.} All experiments were performed on an Intel(R)
Core(TM) i7-4770 3.40GHz/64bit/32GB with Linux 3.13.0 and g++4.8.4. We report
wall-clock times by running each system once and then reporting the average of
four subsequent runs with warm cache. We do not report the times to load the
database into memory for the join as they are orthogonal to this work. All relations are given sorted by their
join attributes.

{\bf Dataset.} We experimented with a real-world dataset in the retail domain
for forecasting user demands and sales. It has five relations: {\tt Inventory}
(storing information about the inventory units for products (sku) in a store
(locn), at a given date), {\tt Census} (storing demographics information per
zipcode such as population, median age, repartition per ethnicities, house units
and how many are occupied, number of children per household, number of males,
females, and families), {\tt Location} (storing the zipcode for each store and
distances to several other stores), {\tt Item} (storing the price and category,
subcategory, and categoryCluster for each products), and {\tt Weather} (storing
weather conditions such as mean temperature, and whether it rains, snows, or
thunders for each store at different dates). The feature extraction query
is the natural join of these five relations. It is acyclic and has 43 variables. We
compute the join over the variable order: {\tt (locn (zip
  ($vars$(Census),$vars$(Location)), date(sku
  ($vars$(Item)),$vars$(Weather))))}.  The following 8 variables are
categorical: {\sf zip, sku, category, subcategory, categoryCluster, snow, rain,
  thunder}. The variables {\sf locn} and {\sf date} are not features in our
models. We design $4$ fragments of our dataset with an increasing number of
categorical features. \textsf{v$_1$} is a partition of the entire dataset that
is specifically tailored to work within the limitations of R. It includes all
categorical variables as features except for {\sf sku} and {\sf
  zip}. \textsf{v$_2$} computes the same model as \textsf{v$_1$} but over all
rows in the data ($5\times$ larger than \textsf{v$_1$}). \textsf{v$_3$} extends
\textsf{v$_2$} with \textsf{zip}, \textsf{v$_2$} and \textsf{v$_3$} are
designed to work within the limitations of MADlib.  \textsf{v$_1$} to
\textsf{v$_3$} have no functional dependency. Finally, \textsf{v$_4$} has all
variables but \textsf{zip} and the functional dependency {\sf
  sku}$\to$\{{\sf category, subcategory, categoryCluster}\}.

We learned $\lr$, $\pr_2$, and $\fama^8_2$ models that predict the amount of
inventory units based on all other features.

  {\bf Summary of findings.} Table~\ref{table:retailer} shows our findings.
  AC/DC+FD is the fastest system in our experiments. It needs up to 30 minutes
  and computes up to 46M aggregates. This is orders of magnitude faster than its
  competitors whenever they do not exceed memory limitation, 24-hour timeout, or
  internal design limitations. The performance gap is due to the optimizations
  of AC/DC: (1) it avoids materializing the join and the export-import step
  between database systems and statistical packages, which take longer than
  computing an end-to-end $\lr$ model in AC/DC. Instead, AC/DC performs the join
  together with the aggregates using one execution plan; (2) it factorizes the
  computation of the aggregates and the underlying join, which comes with a
  20$\times$ compression factor; (3) it massively shares the computation of
  large (up to 46M for $\pr_2$) sets of distinct non-zero aggregates, 
  which makes their computation
  up to 16K$\times$ faster than computing them individually; (5) it decouples
  the computation of the aggregates on the input data from the parameter
  convergence step and thus avoids scanning the join result for each of the up
  to 400 iterations; (6) it avoids the upfront one-hot encoding that comes with
  higher asymptotic complexity and prohibitively large covariance matrices by
  only computing non-identical, non-zero matrix entries. For $\pr_2$ and our
  dataset \textsf{v}$_4$, this leads to a 259$\times$ reduction factor in the
  number of aggregates to compute; (7) it exploits the FD in
  the input data to reduce the number of features of the model, which leads to a
  3.5x improvement factor.

\begin{table*}[t]
\centering
\begin{tabular}{|ll||r|r|r|r|r|}\hline
\multicolumn{2}{|l||}{}    & \texttt{v$_1$} & \texttt{v$_2$}  &  \texttt{v$_3$}  &  \texttt{v$_4$} \\\hline\hline
Join Representation & Listing    & 774M     & 3.614G          & 3.614G           & 3.614G  \\
 (\#values)         & Factorized & 37M      & 169M            & 169M             & 169M \\
 Compression        & Fact/List  & 20.9$\times$ & 21.4$\times$ & 21.4$\times$    & 21.4$\times$ \\\hline
\multicolumn{2}{|l||}{Join Computation (PSQL) for R, TensorFlow, libFM } & 50.63 & 216.56 & 216.56 & 216.56   \\\hline
\multicolumn{2}{|l||}{Factorized Computation of 43 Counts over Join} & 8.02 & 34.15 & 34.15 & 34.15  \\\hline\hline
  \multicolumn{6}{|c|}{\bf Linear regression}\\\hline
Features & without FDs     & 33 + 55 & 33+55 & 33+1340 & 33+3702  \\\cline{3-5}
(continuous+categorical) & with FDs   & \multicolumn{3}{c|}{same as above, there are no FDs} & 33+3653 \\\hline
Aggregates & without FDs   & 595+2,418  & 595+2,421 & 595+111,549 & 595+157,735  \\\cline{3-5}
(scalar+group-by) & with FDs   & \multicolumn{3}{c|}{same as above, there are no FDs} & 595+144,589  \\\hline
MADLib (ols) & Learn     & 1,898.35 & 8,855.11 & $>86,400.00$  & -- \\\hline 
R (QR)     & Export/Import    & 308.83 & -- & -- & -- \\
       & Learn     & 490.13 & -- & -- & --  \\\hline
  TensorFlow (FTLR)          & Export/Import  &  74.72   & 372.70     & 372.70     & 372.70 \\
(1 epoch, batch size 100K)   & Learn          & 2,762.50 & 11,866.37 & 11,808.66 & 11,817.05 \\\hline 
F & Aggregate   & 93.31 & 424.81 & OOM & OOM  \\
         & Converge (runs)   & 0.01 (359) & 0.01 (359) &  & \\\hline
{\bf AC/DC} & Aggregate   & 25.51 & 116.64 & 117.94 & 895.22 \\
         & Converge (runs)   & 0.02 (343) & 0.02 (367) & 0.42 (337)  & 0.66 (365)  \\\hline
{\bf AC/DC+FD} & Aggregate & \multicolumn{3}{c|}{same as {\bf AC}} & 380.31 \\
         & Converge  (runs)    & \multicolumn{3}{c|}{there are no FDs} & 8.82 (366) \\\hline
Speedup of {\bf AC/DC+FD} over & MADlib  & 74.36$\times$ & 75.91$\times$ & $>729.97\times$ & $\infty$ \\
         & R & 33.28$\times$ & $\infty$ & $\infty$ & $\infty$  \\
         & TensorFlow &113.12$\times$ &	106.77$\times$	& 104.75$\times$ & 31.88$\times$  \\
         & F  & 3.65$\times$ & 3.64$\times$ & $\infty$ & $\infty$ \\ \cline{3-5}
         & {\bf AC/DC}  & \multicolumn{3}{c|}{same as {\bf AC/DC}, there are no FDs} & 2.30$\times$ \\\hline\hline
\multicolumn{6}{|c|}{\bf Polynomial regression degree $2$}\\\hline
Features & without FDs     & 562+2,363 &  562+2,366 & 562+110,209  & 562+154,033 \\ \cline{3-5}
(continuous+categorical) & with FDs   & \multicolumn{3}{c|}{same as above, there are no FDs}  & 562+140,936  \\\hline
Aggregates & without FDs  &  158k+742k & 158k+746k & 158k+65,875k & 158k+46,113k  \\ \cline{3-5}
(scalar+group-by) & with FDs & \multicolumn{3}{c|}{same as above, there are no FDs} & 158k+36,712k \\\hline
MADlib (ols)     & Learn     & $>86,400.00$ & $>86,400.00$ & $>86,400.00$ & --  \\\hline 
{\bf AC/DC} & Aggregate     & 131.86 & 512.00 & 820.57 & 7,012.84  \\
         & Converge (runs)   & 2.01 (211) & 2.04 (214) & 208.87 (247) & 115.65 (200) \\\hline 
{\bf AC/DC+FD} & Aggregate & \multicolumn{3}{c|}{same as {\bf AC/DC}} & 1,819.80 \\
         & Converge (runs)    & \multicolumn{3}{c|}{there are no FDs} & 219.51 (180) \\\hline
Speedup of {\bf AC/DC+FD} over & MADlib  & $>645.43\times$ & $>168.08\times$ & $>83,93\times$ & $\infty$  \\ \cline{3-5}
         & {\bf AC/DC}  & \multicolumn{3}{c|}{same as {\bf AC/DC}, there are no FDs} & 3.50$\times$ \\\hline\hline       
\multicolumn{6}{|c|}{\bf Factorization machine degree $2$ rank $8$} \\\hline
Features & without FDs  & 530+2,363 &  530+2,366 & 530+110,209  & 530+154,033  \\\cline{3-5}
(continuous+categorical) & with FDs  & \multicolumn{3}{c|}{same as above, there are no FDs}  & 562+140,936  \\\hline
Aggregates & without FDs         & 140k+740k & 140k+744k &  140k+65,832k & 140k+45,995k  \\ \cline{3-5}
(scalar+group-by) & with FDs & \multicolumn{3}{c|}{same as above, there are no FDs} & 140k+36,595k \\\hline
  libFM (MCMC)  & Export/Import    & 412.84 & 1,462.54 & 3,096.90  & 3,368.06  \\ 
                           & Learn (runs)     & 19,692.90 (300) & $>86,400.00$ (300) & $>86,400.00$ (300) &  $>86,400.00$ (300) \\\hline 
  {\bf AC/DC} & Aggregate  & 128.97 & 498.79 & 772.42 & 6,869.47 \\
                           & Converge (runs)  & 3.03 (300) & 3.05 (300) & 262.54 (300)&  166.60 (300) \\\hline
  {\bf AC/DC+FD} & Aggregate    & \multicolumn{3}{c|}{same as {\bf AC/DC}}  &  1,672.83  \\
                           & Converge (runs)  & \multicolumn{3}{c|}{there are no FDs} & 144.07 (300) \\\hline
  Speedup of {\bf AC/DC+FD} over & libFM  & 152.70$\times$ & >175.51$\times$ &	>86.68$\times$ &  >49.53$\times$\\ \cline{3-5}
                           & {\bf AC/DC}  & \multicolumn{3}{c|}{same as {\bf AC/DC}, there are no FDs} & 3.87$\times$\\\hline
\end{tabular}
\caption{Time performance (seconds) for learning $\lr$, $\pr$, and $\fama$
  models over increasingly larger fragments (\texttt{v$_1$} to \texttt{v$_4$})
  of Retailer. (--) means that the system failed to compute due to design
  limitations. The timeout is set to 24 hours (86,400 seconds). MADlib cannot
  compute any model on \texttt{v$_4$} since the one-hot encoding requires more
  than 1600 columns. R and MADlib do not support $\fama$ models. TensorFlow does
  not support $\pr$ and $\fama$ models. }
\label{table:retailer}
\end{table*}

{\bf Categorical features.} As we move from \textsf{v$_1$}/\textsf{v$_2$} to
\textsf{v$_4$}, we increase the number of categorical features by approx. $50\times$
 for $\lr$ (from 55 to 2.7K) and $65\times$ for $\pr_2$ and $\fama^8_2$ (from
2.4K to 154K). For $\lr$, this increase only led to a $7\times$ decrease in
performance of {\bf AC/DC} and at least $9\times$ for MADlib (we stopped MADlib
after 24 hours). For $\pr_2$, this yields a $13.7\times$ performance decrease
for {\bf AC/DC}. This behavior remains the same for {\bf AC/DC}'s aggregate
computation step with or without the convergence step, since the latter is
dominated by the former by up to three orders of magnitude. This sub-linear
behavior is partly explained by the ability of AC/DC to process many
aggregates much faster in bulk than individually: it takes 34 seconds for 43
count aggregates, one per variable, but only 1819 seconds for 37M sum aggregates!
It is also explained  by the
same-order increase in the number of aggregates: $65\times$ ($51\times$) more distinct
non-zero aggregates in \textsf{v$_4$} vs \textsf{v$_2$} for $\lr$ (resp.\@ $\pr_2$ and $\fama^8_2$).

The performance of TensorFlow is largely invariant to the increase in the number
of categorical features, since its internal mapping from tuples in the training
dataset to the sparse representation of the features vector remains of similar size.
Nevertheless, our system is consistently orders of
  magnitudes faster than computing only {\em a single epoch} in TensorFlow.

{\bf Increasing database size.} A $5\times$ increase in database size and join
result from \textsf{v$_1$} to \textsf{v$_2$} leads to a similar decrease factor
in performance for F and AC/DC on all models, since the number of features and
aggregates stay roughly the same and the join is acyclic and processed in
linear time. The performance of MADlib, TensorFlow, and libFM follows the same
trend for $\lr$ and $\fama$. MADlib runs out of time (24 hours) for both datasets for
$\pr_2$ models. R cannot cope with the size increase due to internal design
limitations.

{\bf One-hot encoding vs.\@ sparse representations with group-by aggregates.}
One-hot encoding categorical features leads to a large number of zero and/or
redundant entries in the $\vec\Sigma$ matrix. For instance, for \textsf{v$_4$} and
$\pr_2$, the number of features is $m=154,595$, and then the upper half of
$\vec\Sigma$ would have $m(m+1)/2 \approx 1.19\times10^{10}$ entries! Most of these
are either zero or repeating. In contrast, AC/DC's sparse representation
only considers 46M non-zero and distinct aggregates. 
The number of aggregates is reduced by $259$x!

Our competitors require the data be one-hot encoded {\em before} learning. The
static one-hot encoding took (in seconds): 28.42 for R on \textsf{v$_1$}; 9.41
for F on \textsf{v$_1$} and \textsf{v$_2$}; 2 for MADlib on \textsf{v$_1$}
to \textsf{v$_3$}; and slightly more than an hour for libFM, due to the
expensive zero-suppression step. TensorFlow one-hot encodes on the fly
during the learning phase and cannot be reported separately.

{\bf Functional dependencies.} The FD in our dataset \textsf{v$_4$} has a
twofold effect on AC/DC (all other systems do not exploit FDs): it effectively
reduces the number of features and aggregates, which leads to better performance
of the in-database precomputation step; yet it requires a more elaborate
convergence step due to the more complex regularizer. For $\lr$, the aggregate
step becomes $2.3\times$ faster, while the convergence step increases
$13\times$. Nevertheless, the convergence step takes at most 2\% of the overall
compute time in this case. For degree-2 models, the FD brings an improvement by
a factor of 3.5$\times$ for $\pr_2$, and 3.87$\times$ for $\fama^8_2$. This is
due to a 10\% decrease in the number of categorical features, which leads to a
20\% decrease in the number of group-by aggregates.

\noindent
{\em Acknowledgments.} 
This project has received funding from the European Union's Horizon 2020
research and innovation programme under grant agreement No 682588. XN is
supported in part by grants NSF CAREER DMS-1351362, NSF CNS-1409303 and the
Margaret and Herman Sokol Faculty Award.

\bibliographystyle{ACM-Reference-Format}
\bibliography{bibtex}

\end{document}